\documentclass[a4paper,11pt]{article}
\usepackage{pos}

\title{GeV-radio correlation in Markarian 421}
\ShortTitle{GeV-radio correlation in Markarian 421}

\author*{Vitalii~Sliusar}
\author{Roland~Walter}
\author{Matteo~Balbo}

\affiliation{Department of Astronomy, University of Geneva,\\ Chemin d'Ecogia 16, CH-1290 Versoix, Switzerland}

\emailAdd{vitalii.sliusar@unige.ch}

\abstract{
Markarian\,421 (Mrk\,421) is a high-synchrotron-peaked blazar showing relentless variability across the electromagnetic spectrum from radio to gamma-rays. We use over 7-years of radio and GeV observations to study the correlation and connected variability in radio and GeV bands. Radio data was obtained in a 15GHz band by the OVRO 40-m radio telescope, and GeV data is from Fermi Large Area Telescope. To determine the location of the gamma-ray emission regions in Mrk\,421 we correlate GeV and radio light curves. We found that GeV light curve varies independently and accurately leads the variations observed in radio. Using a fast-rise-slow-decay profile derived for shock propagation within a conical jet, we manage to reproduce the radio light curve from GeV variations. The profile rise time is comparable with the Fermi-LAT binning the decay time is about 7.6 days. The best-fit value for the response profile also features a 44 days delay between the GeV and radio, which is compatible with the wide lag range obtained from the correlation. Such a delay corresponds to $10^{17}$ cm/c, which is comparable with the apparent light crossing time of the Mrk\,421 radio core. Generally, the observed variability matches the predictions of the leptonic models and suggests that the physical conditions vary in the jet. The emitting region moving downstream the jet, while the environment becomes first transparent to gamma rays and later to the radio.
}

\FullConference{37$^{\rm{th}}$ International Cosmic Ray Conference (ICRC 2021)\\
		July 12th -- 23rd, 2021\\
		Online -- Berlin, Germany}

\begin{document}

\maketitle

\section{Introduction\label{sec:introduction}}

Blazars are a sub-class of active galactic nuclei (AGNs) with a jet pointing close to an observer line of sight \citep{1995PASP..107..803U}. They are classified as radio-loud AGNs, which feature a flat radio spectrum. Blazars are the most numerous population of GeV detected sources by the  Large  Area  Telescope  (LAT)  on  of the Fermi  Gamma-ray Space Telescope \citep{2011ApJ...743..171A}. Due to jet being pointed towards the observer, the radiation produced by the relativistic plasma within the jet experiences Doppler boosting, yielding to such jets outshine the host galaxy. Radiation in all the bands is non-thermal and show no evidence of spectral lines, preventing from direct estimation of the emission region's bulk Lorenz factor. Such estimations, in case of blazars, is usually based on fittings of the spectral energy distribution (SED). The spectral energy distribution is radiation energy dependence on the frequency/wavelength. For blazars it is usually comprised of two distinctive broad humps (components) \citep{1998MNRAS.299..433F}. The first component is commonly extending from radio up to the X-rays, peaking in UV or soft X-rays. The second component may extend up to TeV energies, peaking in the hundreds of MeVs or GeVs. Based on the location of the synchrotron bump blazars may be further subdivided into low-, intermediate- and high-frequency-peaked blazars (LBL, IBL, and HBL, respectively). Blazars are variable in all the energy bands, from radio to the very-high-energy $\gamma$-rays (TeV band). Variability time scales, depending on the source and band, may vary from minutes to months \citep{2015ApJ...811..143P}. Data about the variability in each band and cross-correlation between the bands gives a valuable insight into the jet emission processes, allowing to estimate the emission region size, presence of separate emission components, distinguish between leptonic, hadronic or mixes emission, etc.

Mrk\,421 is a bright blazar located at $z = 0.031$. It belongs to the subclass of high-frequency-peaked blazars with synchrotron bump peaking in the soft X-rays \citep{abdo_2011ApJ...736..131A}. Mrk\,421 is a bright $\gamma$-rays source with frequent periods of flaring activities. The SED of Mrk\,421 can be reproduced using different emission models: e.g.: leptonic one-zone synchrotron self-Compton model (SSC) \citep{abdo_2011ApJ...736..131A}, hadronic \citep{2015MNRAS.448..910C,2017A&A...602A..25Z} or lepto-hadronic model \citep{2017A&A...602A..25Z, 2013MNRAS.434.2684M}. To characterize the SED and study Mrk\,421 variavility in different bands, numerous multi-wavelength (MWL) campaigns were caried out to date, observing Mrk\,421 from radio to TeV $\gamma$-rays \citep[e.g.][]{abdo_2011ApJ...736..131A,2021A&A...647A..88A}. The highest variability of Mrk\,421 was found in the X-rays and TeVs \citep{2021A&A...647A..88A} with compatible to zero lag between the two bands \citep{2021A&A...647A..88A, 2021MNRAS.504.1427A}.

In this paper we investigate the GeV-radio correlation using the same data set as was used in \citep{2021A&A...647A..88A}. The analysis is limited to over 7 years of GeV and radio data, from November 2010 to April 2018. The manuscript is structured as follows. In Sect.~\ref{sec:data} we provide an overview of the data set and data reduction applied. The analysis of the variability and correlations between the GeV and radio bands is presented in Sect~\ref{sec:timing}. Sect~\ref{sec:conclusions} includes a brief discussion of the results and the conclusions.

\section{GeV and radio data\label{sec:data}}

The data set used for this study includes over 7-years long GeV and radio light curves. Light curves span from December 2012 to April 2018. We investigate the correlation and lag between the two bands and attempt to find a direct relation between them. During the period of the observation, Mrk\,421 was found in all possible states, including high state period in 2013 and relative quiescent state in late 2016.

The LAT instrument onboard the Fermi Gamma-ray Space Telescope ({\it Fermi}-LAT) is pair-conversion gamma-ray telescope, sensitive to photons from 20\,MeV to over 300\,GeV. It is the most sensitive $\gamma$-ray telescope to date in this energy band. {\it Fermi}-LAT is equiped with a charged particle tracker and a calorimeter to detect high energy $\gamma$-rays from
wide FOV cone. The point spread function (PSF) is artificially estimated by tracking the direction of the arriving photons. The PSF of {\it Fermi}-LAT significantly depends on the energy of the incoming $\gamma$-rays\citep{2012ApJS..203....4A}, reaching about $1\sigma$-equivalent containment radius of $\sim 0.1^{\circ}$ at 40\,GeV \citep{2009ApJ...697.1071A}. Despite {\it Fermi}-LAT being sensitive to $\gamma$-rays of about 20\,MeV, due to the energy-dependent PSF we can only reliably consider the photons with the energy over 100\,MeV, reaching up to 300\,GeV. To obtain a multi-year light curve of Mrk\,421 the PASS8 pipeline and Fermi Science Tool v10r0p5 package were used. Background was estimated using maximum likelihood approach considering the sources from the {\it Fermi}-LAT 4-year Point Source Catalogue.

As a part of {\it Fermi} blazars monitoring campaign \citep{Richards_2011ApJS..194...29R}, Mrk\,421 is observed regularly by the Owens Valley Radio Observatory (OVRO). The main dish of the observatory is a 40\,m radio telescope sensitive particularly in 15\,Ghz band with 3 GHz bandwidth. The typical thermal noise reaches 4 mJy leading to a $\sim 3\%$ uncertainty. The light curve with twice-per-week observations were publicly available from the OVRO official archive website\footnote{http://www.astro.caltech.edu/ovroblazars/}.

\section{Light curves correlation analysis}
\label{sec:timing}
\begin{figure}[h!]
  \centering
      \includegraphics[width=\columnwidth]{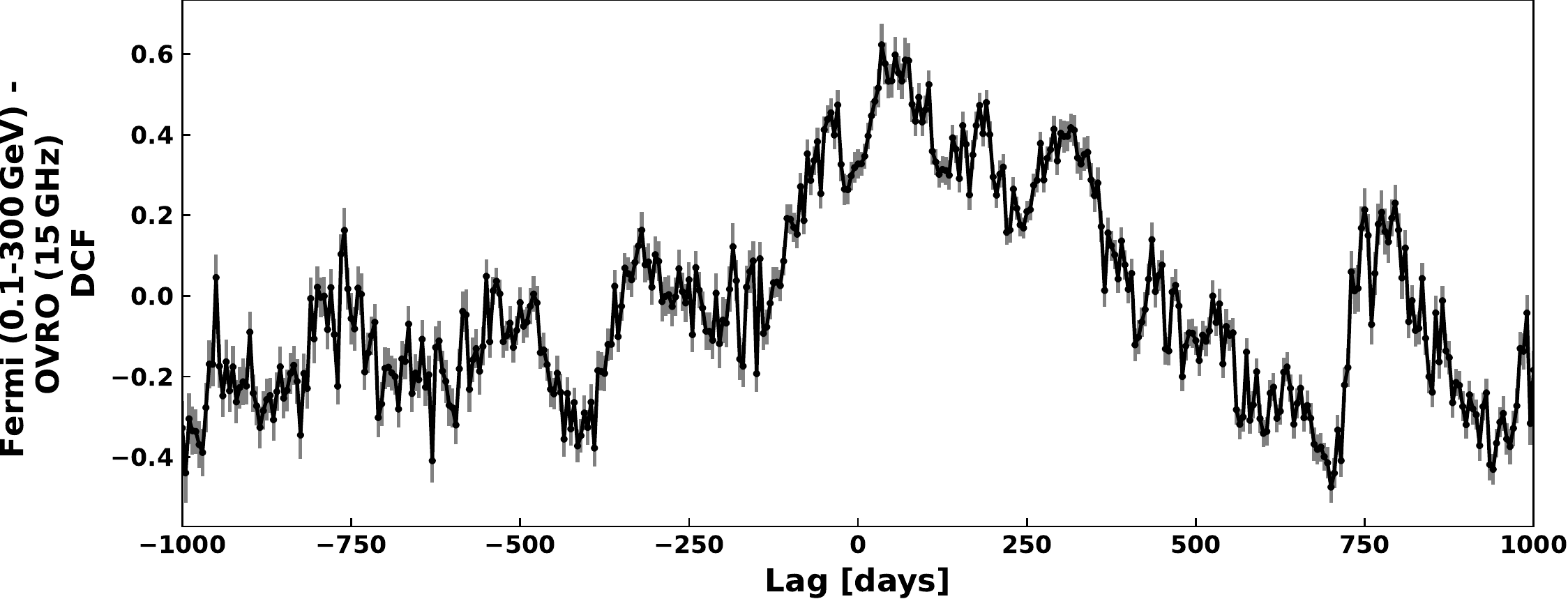}
  \caption{Cross-correlation of the {\it Fermi}-LAT and radio light curves (MJD $\in$ [555000, 58226]). The time resolution is five days. Grey error bars denote 1$\sigma$ uncertainties.}
  \label{fig:cc_fermi_radio}
\end{figure}

The radio light curve is broadly correlated to the GeV light curve with a lag of $30-60$ days at the maximum of the DCF (Fig. \ref{fig:cc_fermi_radio}). The delay of $40\pm9$ days was already reported in \cite{max-moerbeck_2014MNRAS.445..428M} for a specific flaring period of Mrk\,421. Existence of a delay between the GeV and radio variations was interpreted as a propagation of a shock through the jet of Mrk\,421, while the environment initially transparent to the $\gamma$-rays and only further downstream to the radio. Similar interpretation was proposed to explain the long term light curves of 3C 273 \citep{turler_1999A&A...349...45T}. Same approach and profile were used to reproduce the radio flares corresponding to overlapping stretched and delayed GeV flares \citep{esposito_2015A&A...576A.122E}.

\begin{figure}[h]
  \centering
      \includegraphics[width=\columnwidth]{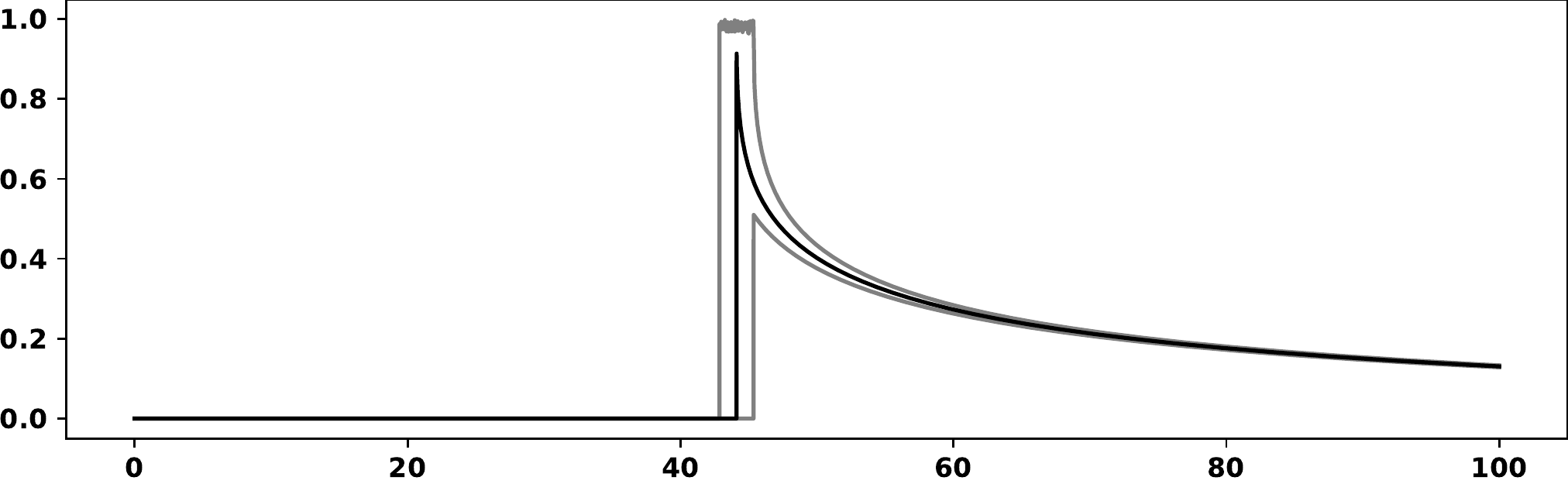}
  \caption{Radio response profile to the GeV light curve, the $y$ axis is in arbitrary units. The grey lines denote the 1$\sigma$ uncertainties on the fitted parameters.}
  \label{fig:radio_profile}
\end{figure}

Using the 5-years long light curves in GeV and radio bands we investigate the possibility to reconstruct the radio light curve of Mrk 421 by convolving the GeV light curve with a special function - a response profile. We assume that the electrons emitting GeV through inverse Compton emission, would eventually also emit delayed (relatively to the GeV band) and somewhat stretched synchrotron flares in the radio band. We adopt the same asymmetric response profile as proposed in \citep{turler_1999A&A...349...45T}. Unlike \citep{turler_1999A&A...349...45T}, where the profile was used for individual flares produced by a shock moving through the conical jet, we try to estimate a function usable through multi-year time range. We found that in case of Mrk\,421, the rise time ($t_{rise}$) is comparable with the {\it Fermi}-LAT binning time of 2 days, so instead of the original profile as in \citep{turler_1999A&A...349...45T}, we adopt a simpler four parameters immediate-rise-slow-decay response profile (see Fig. \ref{fig:radio_profile}):

\begin{equation}
\label{eq:profile_step}
  S(t) =
    \begin{cases}
      0 & \text{$t < \Delta t$}\\
      A  \exp\left(-\left(\frac{t - \Delta t}{t_{decay}}\right)^\phi \right) & \text{$t \geq \Delta t$}\\
    \end{cases}
\end{equation}

\begin{figure}[h!]
  \centering
      \includegraphics[width=\columnwidth]{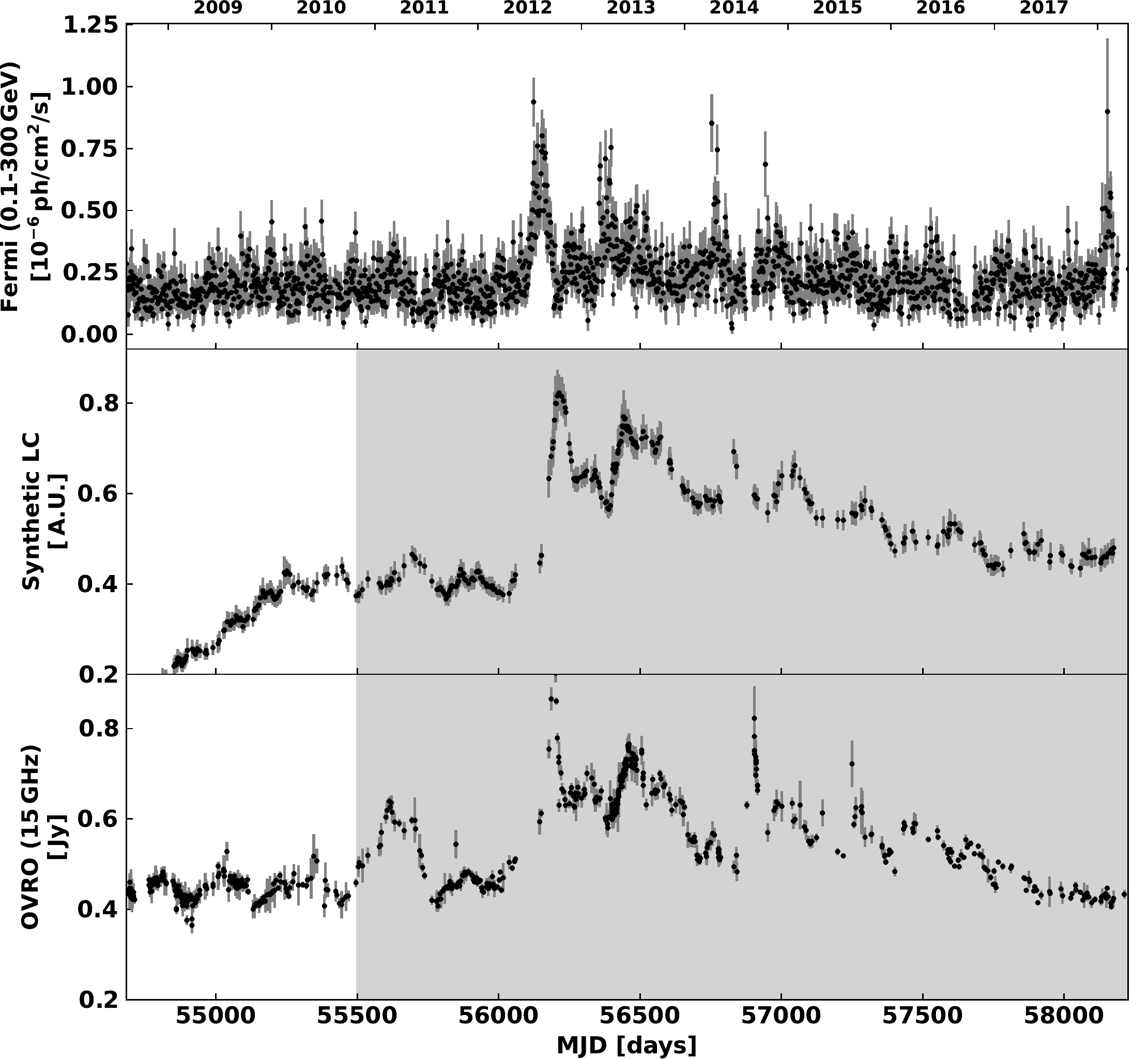}
  \caption{Synthetic radio light curve (middle), created as a convolution of the two-days binned {\it Fermi}-LAT 0.1-300\,GeV light curve (top) and of the radio response (Fig.~\ref{fig:radio_profile}), compared with the OVRO 15\,GHz radio light curve (bottom). The time range highlighted in grey for the radio and synthetic light curve indicate the region considered for the fit of the response profile.}
  \label{fig:fermi_radio_conv}
\end{figure}

The best fit response was obtained by minimizing the deviations between the observed radio light curve and the synthetic one over the 7.5 years period (MJD 55500-58226). The synthetic light curve was obtained by convolving the {\it Fermi}-LAT light curve with a response function (Eq. \ref{eq:profile_step}). Since the {\it Fermi}-LAT light curve has a starting point and the delay has a gradual decay, this yields to gradual increase of synthetic radio light curve flux. Thus the minimization period starts about two years after the {\it Fermi}-LAT light curve to eliminate the the long lasting effects of the response gradual decay. We used a four-parameter response with an exponential decay after a delay ($\Delta t$). Similar outburst profile, but with additional two parameters to describe the rise of the response before decay phase \citep{turler_1999A&A...349...45T}, was used in \cite{esposito_2015A&A...576A.122E}.  The observed GeV, radio and the resulting synthetic radio light curve are shown in Fig. \ref{fig:fermi_radio_conv}, the best fit parameters are listed in Table~\ref{tab:profile}. The uncertainties of the synthetic light curve are derived using direct uncertainty propagation from the {\it Fermi}-LAT light curve through the response function.

\begin{table}
\centering
\caption{Best fit parameters of the $\gamma$-ray ({\it Fermi}-LAT) to radio response profile (see Eq.~\ref{eq:profile_step}) in Mrk\,421.}
\label{tab:profile}
\begin{tabular}[t]{lc}
Parameter & Value \\
\hline
\hline

$A$ & $1.4 \times 10^{5}$ \\
$t_{decay}$ & 7.64 days \\
$\phi$ & 0.36 \\
$\Delta t$ & 44.1 days \\ \hline
\end{tabular}
\end{table}%

The best-fit synthetic light curve is similar, but does not match perfectly the observed radio light curve, possibly indicating that the intensity of the response (the constant A) might be variable with time. Nevertheless with a single response profile we are capable of reproducing the radio light curve for over 7 years. The fast radio flare near MJD 56897 and a wider flare at about MJD 55600 (see Fig. \ref{fig:fermi_radio_conv}) could not be reproduced using a stationary response profile. These flares could have a different origin, or may need a different response (such adjustments were also used by \cite{esposito_2015A&A...576A.122E} and could indicate different conditions in various shocks). As the uncertainties of the GeV light curve and of the response profile are not Gaussian nor independent, the goodness of fit $(\chi_{\nu}^2 = 1.5; \nu=281)$ between the radio data and the convolved GeV light curve is only indicative though used during the minimization. The response decay time is constrained to a small range ($7.64\pm0.11$ days). The uncertainty of $\pm 1.3$ days on the delay $\Delta t$ is similar to the time bin duration of the GeV light curve. We also tried the response profile from \cite{turler_1999A&A...349...45T}, including an exponential rise, and found similar parameters than above with a rise time shorter than 3 days. Because of the $\phi$ parameter, $t_{decay}$ is shorter than expected for a purely exponential cutoff.

\section{Conclusions\label{sec:conclusions}}

The analysis of over 7-years long GeV and radio light curves of Mrk\,421 gives an important insight into the radiative processes in this blazar. Strong correlation and low variability (in comparison with X-rays and TeV) hints at the same processes being responsible for the multi-wavelength emission, while the driver for the X-rays/TeV and radio/GeV emission is different (e.g. changes in the cut-off energy). The radio emission can be reproduced by convolving the GeV light curve with a delayed asymmetric response (a fast-rise-slow-decay response profile with a delay of $\sim$44 days). Since using the same profile we are able to reproduce over 7 years of radio data, while the source is found in variety of states, such an approach reveals an internal properties of the source, and connection between its multi-wavelength radiation.

\vspace{1em}
\noindent\textit{Acknowledgements.} This research has made use of data from the {\it Fermi}-LAT \cite{2009arXiv0912.3621S} and from the OVRO 40-m monitoring program \cite{Richards_2011ApJS..194...29R}, supported by private funding from the California Insitute of Technology and the Max Planck Institute for Radio Astronomy, and by NASA grants NNX08AW31G, NNX11A043G, and NNX14AQ89G and NSF grants AST-0808050 and AST- 1109911.

\bibliographystyle{JHEP}
\setlength{\bibsep}{0.4pt}
\footnotesize{
\bibliography{10_references}
}
\end{document}